\documentclass[aps,prl,twocolumn,superscriptaddress,showpacs,floatfix,preprintnumbers]{revtex4-1}

\usepackage[dvipsnames]{xcolor}      
\definecolor{lcolor}{rgb}{0.5,0,0}
\definecolor{citcolor}{rgb}{0,0,1}
\usepackage[breaklinks,colorlinks,urlcolor=blue,citecolor=citcolor,linkcolor=lcolor,linktoc=all]{hyperref}
\usepackage{color}
\usepackage{graphicx}	
\graphicspath{{./figures/}}
\usepackage[utf8]{inputenc}
\usepackage{amsmath} \usepackage{amssymb}
\usepackage[dvipsnames]{xcolor}      

\allowdisplaybreaks

\newcommand{\muB}{\mu_{\text{B}}}
\newcommand{\Lh}{\Lambda_{\text{h}}}
\newcommand{\mE}{m_{\text{E}}}

\newcommand{\NLO}[1]{N$^{#1}$LO}

\renewcommand{\Ref}{Ref.~}
\newcommand{\Refs}{Refs.~}
\newcommand{\eq}{Eq.~}
\newcommand{\eqs}{Eqs.~}

\begin{document}
\title{Soft interactions in cold quark matter}

\preprint{HIP-2021-9/TH}
\author{Tyler Gorda}
\affiliation{Technische Universit\"{a}t Darmstadt, Department of Physics, D–64289 Darmstadt, Germany}
\affiliation{Helmholtz Research Academy for FAIR, D–64289 Darmstadt, Germany}
\author{Aleksi Kurkela}
\affiliation{Faculty of Science and Technology, University of Stavanger, Stavanger, Norway}
\author{Risto Paatelainen}
\affiliation{Helsinki Institute of Physics and Department of Physics, University of Helsinki, Finland}
\author{Saga S\"appi}
\affiliation{European Centre for Theoretical Studies in Nuclear Physics and Related Areas (ECT*) and Fondazione Bruno Kessler, Italy}
\affiliation{Helsinki Institute of Physics and Department of Physics, University of Helsinki, Finland}
\author{Aleksi Vuorinen}
\affiliation{Helsinki Institute of Physics and Department of Physics, University of Helsinki, Finland}

\begin{abstract}
    Accurate knowledge of the thermodynamic properties of zero-temperature, high-density quark matter plays an integral role in attempts to constrain the behavior of the dense QCD matter found inside neutron-star cores, irrespective of the phase realized inside the stars. 
   In this Letter, we consider the weak-coupling expansion of
    the dense QCD equation of state and compute the next-to-next-to-next-to-leading-order contribution arising from the non-Abelian interactions among long-wavelength, dynamically screened gluonic fields. Accounting for these interactions requires an all-loop resummation, which can be performed using hard-thermal-loop (HTL) kinematic approximations. Concretely, we perform a full two-loop computation using the HTL effective theory, valid for the long-wavelength, or soft, modes.
    We find that the soft sector is well-behaved within cold quark matter, contrary to the case encountered at high temperatures, and find that the new contribution decreases the renormalization-scale dependence of the equation of state at high density.

\end{abstract}

\maketitle

\emph{Introduction}---The equation of state (EOS) of dense deconfined quark matter (QM) can be determined from the theory of strong interactions, quantum chromodynamics (QCD), in terms of a perturbative series in the strong coupling constant $\alpha_s$ \cite{Freedman:1976dm,Freedman:1976ub,Kurkela:2009gj,Gorda:2018gpy}. This weak-coupling expansion plays a significant role in constraining the EOS of neutron-star (NS) matter \cite{Kurkela:2014vha,Annala:2017llu,Most:2018hfd} and, in particular, is a crucial ingredient in attempts to determine the physical phase of matter inside NS cores \cite{Annala:2019puf}.

The last fully determined perturbative order for the EOS of cold (i.e., zero-temperature), unpaired QM originates from a next-to-next-to-leading order (\NLO{2}) calculation by Freedman and McLerran in the late 1970s \cite{Freedman:1976dm,Freedman:1976ub} (see also \cite{Fraga:2001id,Vuorinen:2003fs} for $\overline{\rm MS}$ results), later supplemented by nonzero-quark-mass \cite{Fraga:2004gz,Kurkela:2009gj} and nonzero-temperature effects \cite{Kurkela:2016was}.  To extend this result to \NLO{3}, or $O(\alpha_s^3)$ in the strong coupling, several technical and conceptual challenges must be overcome. Besides the self-evident complexity related to four-loop Feynman graphs, the calculation is sensitive to the dynamical screening of long-wavelength chromoelectric and chromomagnetic fields, which requires going beyond a fixed loop order in the perturbative calculation. Indeed, starting at \NLO{2}, each successive order in $\alpha_s$ necessitates the resummation of an infinite number of diagrams.

In stark contrast to the realm of high temperatures $T$ \cite{Braaten:1995cm,Kajantie:2002wa,Kajantie:2003ax}, the inability to formulate an effective-theory description for the long-wavelength modes and perform the required resummation has been one of the major hurdles preventing an \NLO{3} determination of the cold-QM EOS. Here, we finally report the results from completing this resummation, taking a significant step towards obtaining the full \NLO{3} pressure. We work in the limit of vanishing quark masses, which is a very good approximation at the densities where the weak-coupling expansion displays convergence \cite{Kurkela:2009gj}.

In electromagnetic plasmas, dynamical screening leads, e.g., to the well-known physics of Landau damping and to a nonzero plasma frequency \cite{Landau:1946jc}. Analogous phenomena also occur in strongly interacting matter (for a recent review, see~\cite{Ghiglieri:2020dpq}).  The crucial difference between quantum electrodynamics (QED) and QCD is that the screened modes self-interact only in the latter, leading to uniquely non-Abelian physics. Interaction corrections for the screened modes have been discussed using the hard-thermal-loop (HTL) effective theory in the contexts of, e.g., plasmon damping and frequency \cite{Braaten:1990it,Braaten:1992gd,Schulz:1993gf,Carrington:2008dw}, heavy quark diffusion \cite{CaronHuot:2007gq, CaronHuot:2008uh}, thermal photon production \cite{Ghiglieri:2013gia}, and transport \cite{Ghiglieri:2015ala, Ghiglieri:2018dib} in a high-temperature quark-gluon plasma (QGP), as well as for the neutrino interaction rate in an electroweak plasma \cite{Jackson:2019tnr}. 

Based on observations made at high temperatures, it has been conjectured that soft modes exhibit particularly poor convergence and drive the breakdown of the perturbative series of quantities like the pressure in thermal field theory \cite{Braaten:1995jr,Blaizot:2003iq,Gynther:2007bw}. Indeed, reorganizing the weak-coupling expansion in the soft sector of QCD beyond a minimal resummation has been seen to dramatically improve the convergence of the QGP EOS \cite{Blaizot:2000fc,Laine:2006cp,Mogliacci:2013mca}. This raises the obvious question of whether a similar approach should be taken also in studying high-density QM (see \cite{Kneur:2019tao} for an attempt in this direction).

\begin{figure}
    \vspace{-0.5cm}
    \centering
    \includegraphics[width=.40\textwidth]{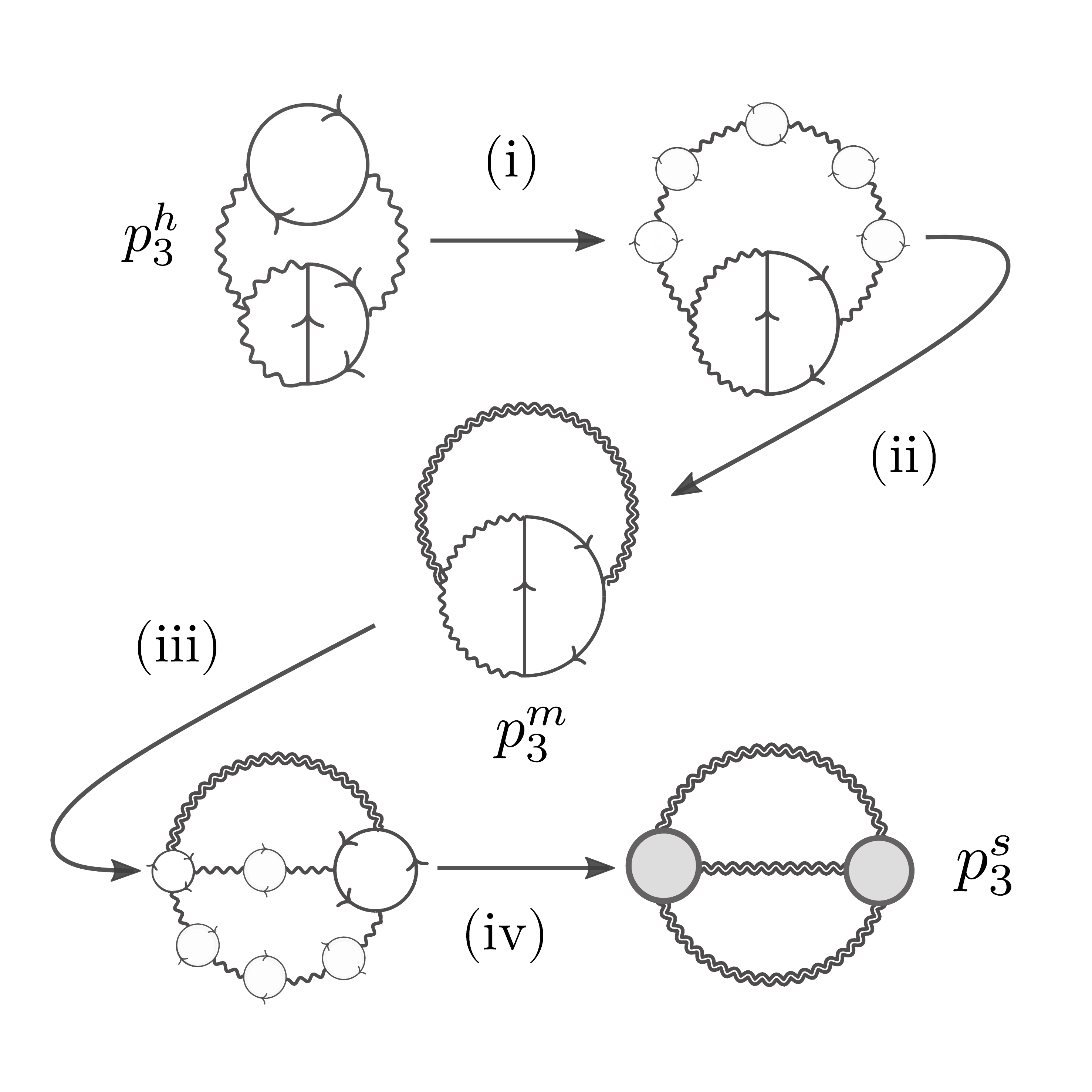}
    \vspace{-0.2cm}
    \caption{An example of how different kinematic regions contribute to the cold-QM EOS at \NLO{3}: (i) When the momentum flowing through a gluon line in a four-loop Feynman graph becomes soft, inserting additional self-energy corrections on the said line does not affect the order of the diagram. (ii) The soft gluon line can be resummed within the HTL theory, producing the effective propagator denoted by a thick wavy line and leading to a three-loop `mixed' contribution (part of $p_3^m$) if the other gluon momenta are hard. (iii) If the other independent gluon momentum in the graph becomes soft, then the other gluon propagators and interaction vertices need to be dressed with additional loops as well, giving rise to (iv) the fully soft HTL contribution $p_3^s$, including resummed HTL vertices.}
    \label{fig:my_label1}
\end{figure}

In the context of cold QM, HTL (or HDL~\cite{Ipp:2003cj,Ipp:2006ij}) methods were introduced already in the 1990s \cite{Manuel:1995td}, and have since been used to study, e.g.,  thermodynamic quantities \cite{Baier:1999db, Andersen:2002jz,Fujimoto:2020tjc} and non-Fermi-liquid behavior  \cite{Ipp:2003cj,Gerhold:2004tb} at the one-loop level. More recently, it was shown in Ref.~\cite{Kurkela:2016was} that the HTL effective theory can be used to systematically resum soft contributions in the weak-coupling expansion of the EOS also at vanishingly small temperatures. In each of these cases, however, the calculations were performed at low enough order that they were not yet sensitive to the interactions between the soft modes. 

In this Letter, we present results from a two-loop HTL calculation that fully accounts for the contributions of the screened field modes and their interactions to the \NLO{3} EOS of high-density, zero-temperature QM. This constitutes the first calculation fully accounting for the interactions between soft gluons in QCD thermodynamics: while multiple works exist discussing two- or three-loop HTL thermodynamics at high temperatures \cite{Andersen:2010wu,Andersen:2011sf,Haque:2014rua}, these works all rely on expansions of the HTL diagrams in powers of thermal masses and thus do not perform the full resummation needed in our case. The details of the calculation presented here are provided in a companion paper \cite{companion}. Additionally there, 
we lay out a road map showing how further improvements to the expansion can be achieved based on our new quantitative understanding of the interplay between the soft and hard field modes.

\emph{Structure of the weak-coupling expansion}---The field modes affected by dynamical screening are soft, with wave number $k \lesssim m_\text{E} $, where $m_\text{E} \sim \alpha_s^{1/2} \mu_q$ stands for the related Debye mass scale and $\mu_q$ for the quark-number chemical potential. At \NLO{3}, the EOS of cold QM possesses the schematic structure 
\begin{align}
    p  = p_\text{FD} 
    +\alpha_s p_1^h  & +\alpha_s^2 p_2^h  + \alpha_s^3 p_3^h \nonumber \\*
    & +\alpha_s^2 p_2^s  +\alpha_s^3 p_3^s  \nonumber  \\*
    &  \quad \quad \quad\, + \alpha_s^3 p_3^m , \label{eq:schematics}
\end{align}
where $p_\text{FD}$ is the pressure of a free Fermi gas of quarks while the other terms are interaction corrections arising from modes of different scales. Terms on the first line arise from hard modes with $k \sim \mu_q$ and can be computed through a naive loop expansion (contributing as $p_i^h \sim \mu_q^4$); terms on the second line arise from soft modes with $k \lesssim \mE$ and their interactions within the HTL theory ($\alpha_s^2 p_i^s \sim \mE^4$);
and the sole term on the third line arises from interactions between the soft and hard modes, requiring a partial HTL resummation ($\alpha_s^3 p_3^m \sim \alpha_s^2 \mE^2 \mu_q^2$).  How these contributions appear from the perspective of full QCD four-loop diagrams is summarized in Fig.~\ref{fig:my_label1} and the corresponding caption. The complete \NLO{3} diagrammatics associated with contributions arising from different momentum regions is also discussed in considerable detail in Sec.~I of \cite{companion}. 

\begin{figure}
\vspace{-0.2cm}
\includegraphics[width=0.45\textwidth]{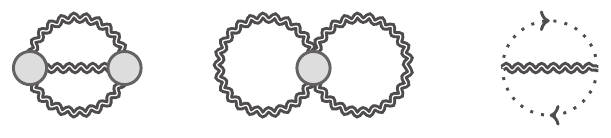}
\vspace{-0.2cm}
    \caption{The two-loop HTL diagrams contributing to the \NLO{3} pressure of cold and dense QM. Note that diagrams with resummed fermionic lines are not needed at any finite order in $\alpha_s$ as long-wavelength fermionic modes are Pauli blocked at zero temperature.}
\label{fig:graphs}
\end{figure}

The above division between soft and hard modes is not unique, as there exists an ambiguity related to the region of semisoft momenta, satisfying $\mE \ll k \ll \mu_q$. Technically, this ambiguity manifests in ultraviolet (UV) divergences in the computation of $p_i^s$ that cancel against corresponding infrared (IR) divergences in $p_i^h$ (as well as mixed UV--IR divergences in $p_3^m$ at \NLO{3}). This renders these coefficients dependent on a factorization scale $\Lh$, which will be canceled in the sum of the different contributions. The precise nature of the factorization scale $\Lh$ depends on the regularization method employed, and it will not in general be a simple cutoff. Combining the terms with logarithmic dependence on $\Lh$ leads then to terms that are enhanced by logarithms of the form $\ln(\mE / \mu_q) \sim \ln(\alpha_s^{1/2})$. Each semisoft loop momentum can produce one such logarithm \cite{Gorda:2018gpy}, so that at the \NLO{3} level we have the decomposition
\begin{align}
    p_2^h + p_2^s &= p_2^\text{LL} \ln \alpha_s +  p_2^\text{const}, \label{eq:logs} \\ 
    p_3^h + p_3^s + p_3^m & = p_3^\text{LL} \ln^2 \alpha_s + p_3^\text{NLL} \ln \alpha_s +  p_3^\text{const}, \nonumber
\end{align}
where the acronyms `LL' and `NLL' refer to the leading and next-to-leading logarithms, respectively. Of these coefficients,  $p_3^\text{LL}$ was originally determined in Ref.~\cite{Gorda:2018gpy} although note that due to a simple computational error the sign of the coefficient reported there is incorrect \footnote{We thank J.-L.~Kneur for bringing this issue to our attention.}. 

In this Letter, we determine the coefficient $p_3^s$ in the above.
This contribution is obtained by summing the three HTL diagrams displayed in Fig.~\ref{fig:graphs}, and constitutes the only term up to \NLO{3} where the self-interactions of soft modes appear.  While $\alpha_s^3 p_3^s$ does not amount to a full order in the weak coupling expansion of the EOS --- for that one would need both the $\alpha_s^3 p_3^h$ and $\alpha_s^3 p_3^m$ coefficients --- it is a physically well-defined and distinct contribution to the quantity, which at least at high temperatures plays an important role in limiting the convergence of the perturbative series \cite{Blaizot:2003iq}.

We leave the technical details of our calculation to a companion paper \cite{companion}. In short, the computation starts by analytically separating the divergent contributions of the diagrams in Fig.~\ref{fig:graphs} from their finite parts.
The divergent terms are found to be generally more amenable to analytic treatment, while multidimensional numerics are required mainly in the evaluation of the finite parts. The technical details of several different parts of the calculation are quite involved and necessitate the development of several new methods, which are explained in full in \Ref\cite{companion}, including its extensive appendices. Here, however, we proceed directly to analyzing the structure and properties of the final result of the computation.

\begin{table}
\begin{tabular}{@{\quad}l@{\quad\quad\quad}l@{\quad}}
\botrule
   $p^{\rm NNLO}_{-1} $ & 1   \\
    $p^{\rm NNLO}_{0}$  & 1.17201  \\
    $p_{-2}$  &  $11/(6 \pi)$ \\
    $p_{-1}$  &   1.50731(19)   \\
    $p_{0}$  &   2.2125(9) \\
\botrule
\end{tabular}
    \caption{List of numerical values for the coefficients in \eqs\eqref{NLLOresult} and \eqref{p}. Note that our normalization conventions for the constants here differ slightly from those used in the companion paper \cite{companion}. }
    \label{tab:my_label}
\end{table}

\emph{Results}---We work consistently in dimensional regularization in $d=3-2\epsilon$ dimensions, wherein the factorization scale $\Lh$ mentioned above \eq\eqref{eq:logs} appears through the integration measure $[e^{\gamma_{\text{E}}}\Lh^2/(4 \pi)]^{\epsilon} {\mathrm d}^{4-2\epsilon}K$, where $\gamma_\text{E}$ is the Euler--Mascheroni constant. The contributions $p_2^s$ and $p_3^s$ in \eq\eqref{eq:schematics} then obtain the forms
\begin{align}
\alpha_s^2 p_2^s & = \frac{d_A \mE^4}{ (8\pi )^2}  \left( \frac{\mE}{\Lh} \right)^{\!\!-2\epsilon} \left( \frac{p^{\rm NNLO}_{-1}}{2\epsilon} + p^{\rm NNLO}_0 \right), \label{NLLOresult} \\
\alpha_s^3 p_3^s &= 
        \frac{\alpha_s N_c d_A \mE^{4}}{ (8\pi)^{2}}  \!
        \left( \frac{\mE}{\Lh} \right)^{\!\!-4\epsilon} \!
        \left( \frac{p_{-2}}{(2 \epsilon)^{2}} + \frac{p_{-1}}{2 \epsilon} + p_{0} \right), 
\label{p}
\end{align}
where the former term originates from the leading-order HTL pressure and the latter from the graphs in Fig.~\ref{fig:graphs}. Here, $N_c$ stands for the number of colors and $d_A = N^2_c - 1$  for the number of gluons associated with SU($N_c$). Finally, the $T=0$ screening mass, reproduced here for the case of multiple massless quark flavors with chemical potentials $\mu_f$, reads $\mE^2 = (2\alpha_s / \pi) \sum_f \mu_f^2$.

\begin{figure}
\centering
\vspace{-0.2cm}
\includegraphics[width=0.49\textwidth]{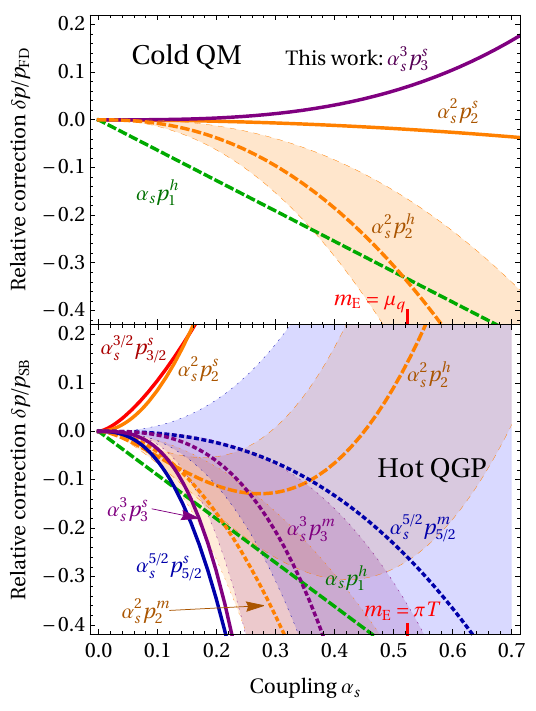}
\vspace{-0.4cm}
    \caption{Relative sizes of different contributions to the pressures of cold QM and hot QGP \cite{Kajantie:2002wa,Laine:2005ai,Ghisoiu:2015uza}. The bands around the $\alpha_s^2 p_2^h$, $\alpha_s^2 p_2^m$, $\alpha_s^{5/2} p_{5/2}^m$, and $\alpha_s^3 p_3^m$ terms arise from varying the renormalization scale $\bar \Lambda = \{1,2,4\} \times \mu_q$ (for cold QM) and $\bar \Lambda = \{1,2,4\} \times \pi T$ (for hot QGP). The mixed terms $\alpha_s^{i} p_{i}^m$ in the high-$T$ expansion are defined as those originating from (hard) corrections to the matching parameters appearing in the soft terms. Note that the contributions associated with the ultrasoft scale $\alpha_s T$ present at nonzero $T$ are not shown in the figure. The factorization scale between the soft and hard sectors is fixed to the cold-QM PMS value, $\Lh^{\rm PMS} \approx 0.275 \mE$, in both cases. On the axes, we show the values of $\alpha_s$ corresponding to where $\mE = \pi T$ and $\mE = \mu_q = \muB/3$, marking the points at which the hierarchy between the hard and soft scales becomes inverted.}
\label{fig:regions}
\end{figure}

\begin{figure}
    \centering
    \includegraphics[width=0.47\textwidth]{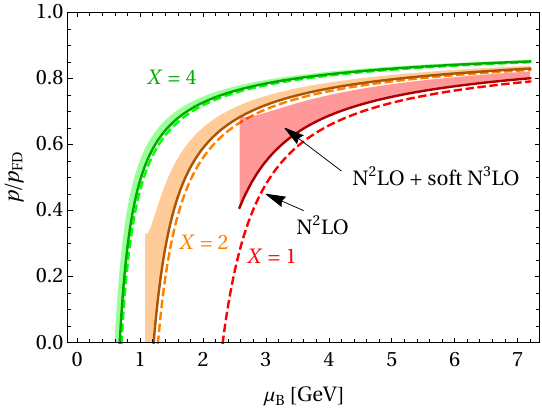}
    \caption{%
    A comparison between the state-of-the-art \NLO{3} EOS of cold QM and the corresponding \NLO{2} result, given as functions of the baryon chemical potential $\muB$, 
    The three colors correspond to the renormalization-scale variation in the hard sector with $\bar \Lambda = X \muB/3$. The solid lines in the \NLO{3} result correspond to fixing the factorization scale $\Lh$ to the PMS scale, while the filled bands result from varying $\Lh$ by a factor of two in both directions about the PMS scale. 
    }
    \label{fig:pmu}
\end{figure}

The values of all coefficients appearing in the above results are given in Table \ref{tab:my_label}. The value of $p_{-2}$ dictates the value of the coefficient $p^{\text{LL}}_3$, first determined in \Ref\cite{Gorda:2018gpy} using a different regularization scheme; this correspondence is explained in detail in Appendix E of the companion paper \cite{companion}. An interesting detail visible in Eqs.~(\ref{NLLOresult}) and (\ref{p}) is that both the \NLO{2} and \NLO{3} expressions depend on the quark chemical potentials only through the parameter $\mE$. This is a direct consequence of the HTL kinematics wherein all the momenta $k$ flowing in the graphs are assumed parametrically smaller than the chemical potentials $\mu_f$. In applying the dimensional-regularization scheme, however, we extend the integration region to infinity, leading to the UV divergences $1/\epsilon^2$ and $1/\epsilon$ in the results. These terms are not true UV divergences of the underlying quantum field theory, eventually canceled by renormalization, but appear because of the kinematic approximations made. They will, by construction, 
cancel in the manner described above \eq\eqref{eq:logs}.

To extract the finite \NLO{3} contribution to the pressure, \eq\eqref{p} can be expanded in powers of $\epsilon$ to yield
\begin{align}
  \alpha_s^3  p_3^s &= \frac{\alpha_{s} N_c d_A \mE^{4}}{(8\pi)^{2}}
    \bigg(
        \frac{p_{-2}}{4 \epsilon^2}
        +
        \frac{p_{-1} - 2 p_{-2}\ln(\mE/\Lh) }{2 \epsilon}
    \nonumber \\
    &+  p_0 - 2p_{-1} \ln(\mE/\Lh) + 2 p_{-2} \ln^2(\mE/\Lh) \bigg).
    \label{pexp}
\end{align}
The terms on the second line constitute our final result, i.e.~the soft $O(\alpha_s^3)$ contribution to the EOS of cold and dense QM.

The convergence properties of the perturbative series can be analyzed by studying the relative magnitudes of the finite $O(\epsilon^0)$ coefficients in the weak-coupling expansion of the pressure, including further splitting each loop order into the contributions from different momentum scales. For simplicity, we restrict ourselves here to $\beta$-equilibrated three-flavor QM, corresponding to equal chemical potentials for the (here massless) up, down, and strange quark flavors, $\mu_f = \mu_q =  \muB/3$, where $\muB$ is the baryon chemical potential. In Fig.~\ref{fig:regions} (top), we show the different contributions as functions of the coupling $\alpha_s$. To minimize the ambiguity in the division of modes to the hard and soft sectors, we have fixed the value of the factorization scale $\Lh$ using the principle of minimal sensitivity (PMS) \cite{Stevenson:1982qw}, $\mathrm{d}p/\mathrm{d}\Lh = 0$, leading to $\Lh^{\rm PMS} = \exp[{-p_{-1}/(2p_{-2}) }] \mE$. The hard contributions also depend on the $\overline{\text{MS}}$ renormalization scale $\bar \Lambda$, which we vary over the range $\bar\Lambda = \{ 1,2, 4 \}\times\mu_q$.

For moderate couplings $\alpha_s \lesssim 0.4$, the dominant correction to the pressure is the NLO term $\alpha_s p_1^h$ arising from hard modes. At larger couplings, the \NLO{2} hard correction $\alpha_s^2 p_2^h$ begins to dominate, signaling the breakdown of the perturbative series. At these values of the coupling, the leading-order contribution from soft modes $\alpha_s^2 p_2^s$ is very small, and while the soft interaction correction $\alpha_s^3 p_3^s$ is significantly larger, it still remains subdominant to the hard corrections. Therefore our results demonstrate that soft contributions do not always drive the breakdown of the weak-coupling expansion, contrary to the standard lore. In fact, $\alpha_s^3 p_3^s$ is the first positive correction in the perturbative series, thus somewhat improving the convergence of the expansion. 

It is very interesting to contrast these cold-QM results to earlier hot-QGP results. As the lower panel of Fig.~\ref{fig:regions} demonstrates, the converge of the pressure expansion is vastly inferior at high temperatures, which has in fact driven the extensive development of resummation schemes based on the HTL \cite{Andersen:1999fw,Andersen:1999sf} and electrostatic-QCD \cite{Laine:2006cp} effective theories. Another qualitative difference seen in Fig.~\ref{fig:regions} is that the soft corrections at high-$T$ are significantly larger than the hard ones, in stark contrast to the high-density results.

Using the two-loop running of the strong coupling $\alpha_s(\bar \Lambda)$, we finally present in Fig.~\ref{fig:pmu} the full QM pressure as a function of $\muB$. This  represents the state-of-the-art perturbative result for the quantity, superseding earlier works such as \Refs\cite{Freedman:1976dm,Freedman:1976ub,Kurkela:2009gj,Gorda:2018gpy}. In addition to including scale variation in  $\bar \Lambda$, we additionally display the variation of $\Lh$, to be cancelled upon the determination of the $\alpha_s^3 p_3^m$ and $\alpha_s^3 p_3^h$ terms, as filled bands representing the choices $\Lh= \{1/2,1,2\} \times \Lh^{\text{PMS}}$. The solid lower line corresponds to the central value $\Lh = \Lh^{\text{PMS}}$. We note that for fixed $\muB$, the pressure is always somewhat higher and  tighter constrained than the corresponding \NLO{2} result. In this context, it is important to note that a straightforward analytical calculation shows that the PMS value for $\Lh$ also maximizes the scale variation with $\bar \Lambda$ of our new result. 

\emph{Conclusions and discussion}---In this Letter, we have reported results from a new state-of-the-art calculation of the EOS of cold and dense QM that treats the soft screened modes to \NLO{3} accuracy. This places the zero-temperature result nearly on par with its high-temperature counterpart, which is known up to but not including the hard contribution. From Fig.~\ref{fig:pmu}, we observe that in stark contrast to the high-temperature case, the perturbative terms from the soft sector appear to converge rather well for cold QM. This can be attributed to two crucial differences between dynamical screening in cold and hot matter. First, due to the Bose enhancement of long-wavelength modes at nonzero temperature, the soft sector is expanded in powers of $g = \sqrt{4 \pi \alpha_s}$, rather than in integer powers of $\alpha_s$. Additionally, the coefficients of the expansion are seen to take significantly smaller values in cold QM, perhaps due to the more moderate screening effects. This suggests that perturbation theory may be much more powerful for extracting the properties of cold QM than it is for hot QGP.  We furthermore emphasize that our results are independent of the many color-superconducting phases that may be present in cold QM since only gluons with nonpertubatively small energies and momenta $\sim \exp(-\# / g)$ receive corrections to their screening from quark pairing \cite{Malekzadeh:2006ud,Alford:2007xm}. Hence, no changes to the EoS  appear at any finite order in $\alpha_s$.

As always in high-order loop calculations, one may observe novel and interesting structures emerge. Here, one such structure arises from the analytical computation of $p_{-2}$ and $p_3^{\rm LL}$, which we find to be closely related to the $\beta$ function of pure gauge theory. We speculate that this must be so because the same semisoft gluonic modes give rise to both the UV divergence of the (pure-glue) HTL theory and to the leading-logarithmic contribution to the pressure \footnote{Note that in QED, the pure-gauge part of the $\beta$ function vanishes. This is consistent with the fact that pure-gauge HTL diagrams are absent in this case, since the HTL effective vertex between $N \ge 3$ photons vanishes \cite{Bellac:2011kqa}.}. It may be that this connection can be used to determine the leading-logarithmic contributions at higher orders, but we leave a detailed inspection of this issue for future work. 

Lastly, we note that our new soft contribution $\alpha_s^3 p_3^s$ pushes the pressure to higher values at a given fixed $\muB$, thus reducing the scale-variation error and pushing the result more towards the pressure of a gas of free quarks.  We note that this is consistent with the picture emerging from an empirical determination of the EOS using astrophysical observations together with perturbative-QCD and nuclear-theory calculations \cite{Annala:2017llu,Annala:2019puf}. In these works it has been repeatedly seen that the EOS continues a trend set by the perturbative calculation even at densities where this computation displays large uncertainties related to the truncation of the weak-coupling expansion. Whether the increasing trend continues upon the determination of the next unknown term in the expansion, $\alpha_s^3 p_3^m$, will be very interesting to investigate.

\emph{Acknowledgements}---RP, SS, and AV have been supported by the Academy of Finland grant no.~1322507, as well as by the European Research Council, grant no.~725369. This work was begun while TG was supported by the U.S.\ Department of Energy under Grant No.\ DE-SC0007984.

\bibliography{References.bib}

\end{document}